\documentclass[%
 reprint,
superscriptaddress,
 amsmath,amssymb,
 aps,
]{revtex4-2}

\usepackage{graphicx}
\usepackage{dcolumn}
\usepackage{bm}
\usepackage{hyperref}
\hypersetup{
     colorlinks=true,
     linkcolor=blue,
     filecolor=magenta,
     urlcolor=blue,
     citecolor=blue,
     }

\begin{document}

\preprint{APS/123-QED}

\title{Anomalous transport regime in non-Hermitian,\\ Anderson-localizing hybrid systems}

\author{Himadri Sahoo}
\affiliation{Nano-optics and Mesoscopic Optics Laboratory, Tata Institute of Fundamental Research, 1 Homi Bhabha Road, Mumbai 400005, India}
\affiliation{Nanophotonics Laboratory, Department of Physics and Nanotechnology, SRM Institute of Science and Technology, Kattankulathur, Tamil Nadu 603203, India}
\author{R Vijay}%
\affiliation{Department of Condensed Matter Physics and Materials Science, Tata Institute of Fundamental Research, 1 Homi Bhabha Road, Mumbai 400005, India}%
\author{Sushil Mujumdar}%
 \email{mujumdar@tifr.res.in}
\affiliation{Nano-optics and Mesoscopic Optics Laboratory, Tata Institute of Fundamental Research, 1 Homi Bhabha Road, Mumbai 400005, India}%

\date{\today}

\begin{abstract}
In a disordered environment, the probability of transmission of a wave reduces with increasing disorder, the ultimate limit of which is the near-zero transmission due to Anderson localization. Under localizing conditions, transport is arrested because the wave is trapped in the bulk of the sample with decaying-exponential coupling to the boundaries. Any further increase in disorder  does not modify the overall transport properties. Here, we report the experimental demonstration of a hitherto-unrealized anomalous transport of hybrid particles under localizing disorder in a non-Hermitian setting. We create hybrid polariton-photon states in a one-dimensional copper sample with a comb-shaped periodic microstructure designed for microwave frequencies. Metallic dissipation realizes the necessary non-Hermiticity. Disorder is introduced by deliberate alterations of the periodic microstructure. Direct measurement of wave-functions and phases was achieved by a near-field probe. At a particular disorder, We observe the onset of Anderson localization of the hybrid states endorsed by exponential tails of the wavefunction. However, at stronger disorder and under conditions that support localization, an unexpected enhancement in the transmission was facilitated by an emergent mini-band. The transmission was traced to the hopping of the hybrid particle over multiple co-existing localized resonances that exchange energy due to the non-orthogonality. These emergent states are manifested in all configurations under strong disorder, suggesting the formation of a novel transport regime. This is verified by measuring the averaged conductance which endorses an anomalous transport regime in the hybrid, non-Hermitian environment under strong disorder. These experimental observations open up new unexplored avenues in the ambit of disorder under non-Hermitian conditions.
\end{abstract}

\maketitle


One of the most exotic phenomena in mesoscopic physics of disorder is Anderson localization (AL), which is synonymous to quasiparticle trapping facilitated by destructive self-interference of quantum waves associated with the quasiparticle. AL occurs at the limit of strongest disorder, with subsidiary regimes of transport such as weakly localized, diffusive, quasi-ballistic and ballistic manifesting with systematic reduction in disorder. Regardless of whether the quasiparticle is electronic\cite{anderson1958absence,pendry1987quasi}, photonic\cite{wiersma1997localization,segev2013anderson,garcia2012nonuniversal,wiersma2013disordered,chabanov2000statistical, smolka2010quantum,pandey2017direct,kumar2020discrepant,mondal2019optical}, phononic\cite{hu2008localization} or of quantum matter\cite{billy2008direct}, the degree of disorder places the transport in one of the above regimes. Of late, however, novel and unexpected behavior has been reported from non-Hermitian photonic systems that belies the underlying strength of disorder\cite{balasubrahmaniyam2018anderson,balasubrahmaniyam2020necklace}. 

The advent of deliberate non-Hermiticity has led to striking developments in photonics in recent years.  Engineered non-Hermitian structures have revealed a plethora of exciting photonic behaviors, such as the striking revelation of exceptional points associated with P-T symmetry implemented through simultaneous loss andgain\cite{feng2017non,longhi2018parity,miri2019exceptional}. In the domain of complex systems, random lasers were the first non-Hermitian systems\cite{wiersma2008physics,cao1999random,cao2005review,mujumdar2004amplified}, although non-Hermiticity was not invoked in the physics thereof. Deliberate non-Hermiticity introduced theoretically in topological disordered lattices exhibited a gapless energy spectrum in the metallic phase, in stark contrast to the gapped spectrum in the Hermitian case\cite{zeuner2015observation}. Furthermore, in two-dimensional lattices, non-Hermiticity was shown to modify the level-spacing statistics\cite{tzortzakakis2020non}, and in fact for three-dimensional lattices, resulted in converting the Anderson transition into a cross-over\cite{huang2020anderson}. Apart from inclusion of gain (negative imaginary refractive index), systems with pure dissipation  also exhibit unusual transport behavior. For instance, measurements on diffusive transport in open systems\cite{wang2011transport,davy2018selectively} helped in identifying specific effects caused by non-Hermiticity, namely, increased eigenfunction correlation accompanied by strong modal anticorrelation that facilitates normalized transport\cite{davy2019probing}. Further, the presence of non-Hermiticity erases the equivalence among total excited energy, dwell time and density of states(DoS), that otherwise exists in the Hermitian system\cite{huang2021wave}. Recently, a surprising development was reported in which a localized system was shown to exhibit transport by a novel mechanism through sudden jumps between distant sites\cite{tzortzakakis2021transport}, a result which has also been experimentally verified\cite{weidemann2020non,weidemann2020nonhermitian}. Overall, one can expect non-Hermiticity to continue unravelling novel effects in quasiparticle transport in mesoscopic systems.

On a parallel note, ongoing studies in mesoscopic transport are exploring an entity of particular interest, namely, composite or hybrid quasiparticles. These particles are created in systems wherein strong interactions between two forms of energy are induced by the underlying structure. They react to a disordered environment with novel emergent behavior. For instance, novel topological phases were discovered in photon-phonon composites in cavity optomechanical systems\cite{peano2015topological}, which were further shown to exhibit a nontrivial frequency dependence of localization length in a disordered environment\cite{roque2017anderson}. Interestingly, Anderson localization was demonstrated to be a potential resource to enhance the coupling between phononic and photonic components to manifest  strongly hybridized composites\cite{garcia2017optomechanical}. The challenges in colocalization of the two components has been overcome in cleverly-designed GaAs/AlAs random superlattices\cite{arregui2019anderson}. 

Given the intricacies that non-Hermiticity and hybridization bring to mesoscopic transport, one can expect exciting consequences of simultaneous presence of both. Precisely such a scenario is addressed in this paper. We experimentally unravel a novel anomalous transport regime at localizing disorder in a hybrid non-Hermitian system. Spoof surface plasmon polaritons (SSPPs)\cite{shen2013conformal} realized in a corrugated metal structure were employed as the composite particles. The SSPPs are generated due to the hybridization of the cavity resonant mode with the polariton mode. We have confirmed the occurrence of the anomalous transport regime (ATR) by measuring the generalized conductance in the disordered system. In corroboration of our earlier theoretical work, we find that the transport is effected via the manifestation of necklace states\cite{chen2011occurrence,chen2013characterization,ghulinyan2007formation,bertolotti2005optical,bertolotti2006wave,wang2020level,choi2015excitation,sgrignuoli2015necklace,bliokh2008coupling}, as identified from direct measurements of phase-jumps occurring in an emergent miniband at high disorder. Concomitant simulations reveal the constituent localized resonances in the measured necklace states. A statistical study of the necklace states reveals the high-probability of their occurrence, and hence the certainty of the ATR in such systems.

\begin{figure}
\centering
\includegraphics[width=1\linewidth]{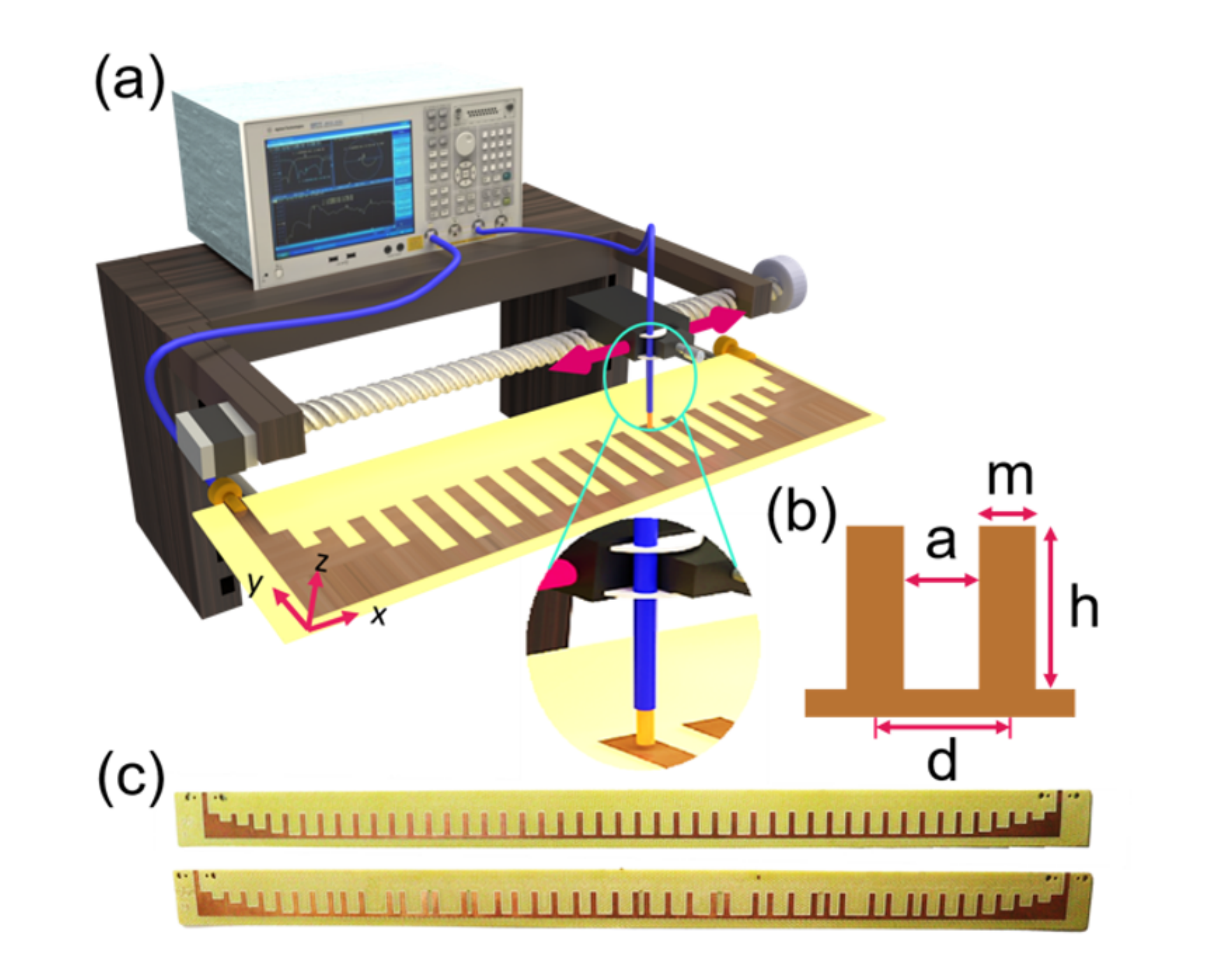}
\caption{Experimental setup and SSPP structure. (a) Schematic of the near-field microwave experimental setup comprising the SSPP sample and a vector network analyzer for source and detection. (b) Unit cell of the SSPP structure with dimensions d~=~6~mm (lattice constant); m~=~2~mm (metal corrugation width);  h~=~7~mm (height) and t=35$\mu$m (metal film thickness). (c) Images of the fabricated samples using standard PCB fabrication technique: periodic (top) and disordered (bottom)}\label{fig:setup}
\end{figure}

Our experimental samples comprise the well-established spoof surface plasmon polariton (SSPP) structures in the form of 1D arrays of microwave resonators cast into comb shaped corrugated metal strips as shown in Fig.\ref{fig:setup}(a). In a single unit cell of periodicity $d=6$~mm, the groove depth and width are denoted by $h$ and $a$, respectively, and the strip width and thickness are $m$ and $t$, respectively fig.\ref{fig:setup}(b). The structure is sculpted, using a commercial circuit board milling machine, onto an FR4 dielectric substrate (a commercial printed circuit board (PCB)) with $\epsilon_r = 4.4$ and total thickness of 1.45~mm. Several periodic samples of metallic corrugated strips with varying $h$ were fabricated using the milling machine and experimentally tested at designated microwave frequencies. Disorder is introduced into the system by randomly displacing the position of each resonator by an amount determined by a uniformly distributed random variate $\Sigma  \in [-\delta \cdot a/2, \delta \cdot a/2]$, where $0\le \delta \le 1$ represents the disorder strength. Multiple configurations of disordered samples with $\delta = 0.3, 0.5$ (referred to as moderate disorder, hereafter) and $0.7$ and $0.9$ (strong disorder), were fabricated. Fig.\ref{fig:setup}(c) shows a fabricated periodic and a disordered sample. We built impedance-matching structures at the output and input ends of the array to ensure efficient launch of the microwave signal into the sample. This was accomplished by constructing metal teeth with gradually increasing heights over a few unit cells at the two ends\cite{gao2014ultra}. The experimental setup consisted of an Agilent E5071C Vector Network Analyzer (VNA), an SMA connector to input the microwave signal at one end and a monopole antenna that acts as the detector. The other end was terminated with a $50~\Omega$ resistor. The detector, attached onto an XY translation stage, is numerically controlled by a stepper motor. The sample was mounted taut horizontally such that the probe distance from the surface of the sample remained $\sim1.5$~mm at any point along the X axis (axis of translation). The scanning step was 1 mm, to obtain high-resolution images. The probe picked up the complex $E_z$ field during the scan, which was recorded by the VNA. In support of the experimental measurements, finite element computations were carried out using the RF module of COMSOL Multiphysics software. Eigenmode analysis was carried out in the simulation for the same structures as in the experiments, as well as transmission profiles were calculated by implementing a lumped port as a source of excitation.

 \begin{figure*}[ht]
\centering
\includegraphics[width=1\linewidth]{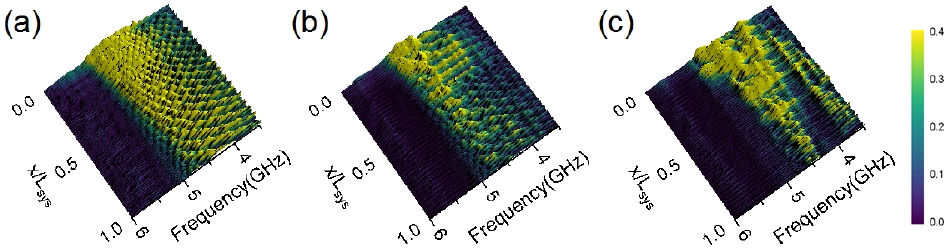}
\caption{Measured spatial intensity distribution $\textbf{I(x)}$ maps for three structures.(a) periodic ($\delta = 0.0$), (b)moderate disorder ($\delta = 0.3$) and (c)high disorders ($\delta = 0.7$). A high-transmission mode close to $4.5$~GHz seen in (c).}
\label{fig:2dfield}
\end{figure*}

The experimentally measured spatial intensity distribution, $I(x)$, is shown in Fig.\ref{fig:2dfield}. Image \ref{fig:2dfield}(a) represents the intensity in the periodic structure, while \ref{fig:2dfield}(b) and \ref{fig:2dfield}(c) show the same for two particular configurations at moderate disorder ($\delta =  0.3$) and high disorder($\delta = 0.7$) respectively. The sample is excited from one end at $x/L_{sys} = 0$, where $L_{sys}$ is the total system length. The spectrally-resolved field is recorded at each $x$-position, at which the source signal is swept through a frequency range of 3.5~GHz to 6~GHz. For the periodic system (\ref{fig:2dfield}(a)), extended modes are formed across the system for all frequencies upto the bandedge at 4.77~GHz. The bandgap (beyond 4.77~GHz) is completely devoid of modes. These frequencies were in excellent agreement with the finite-element calculations, as elaborated in Supplementary Information S1. Images \ref{fig:2dfield}(b) and \ref{fig:2dfield}(c) depict the behavior at intermediate and high disorder, namely $\delta = 0.3~\&~ 0.7$ respectively. Beyond 4~GHz, the transport decays substantially due to the strong disorder. Nonetheless, extended modes are observed around the frequency of 4.5~GHz despite the strong disorder, leading to high transmission. That these extended modes were of the character of necklace states was directly inferred from two diagnostics, namely, the experimentally-measured phase characteristics and the computed eigenmode analysis.

\begin{figure}[h]
\centering
\includegraphics[width=1\linewidth]{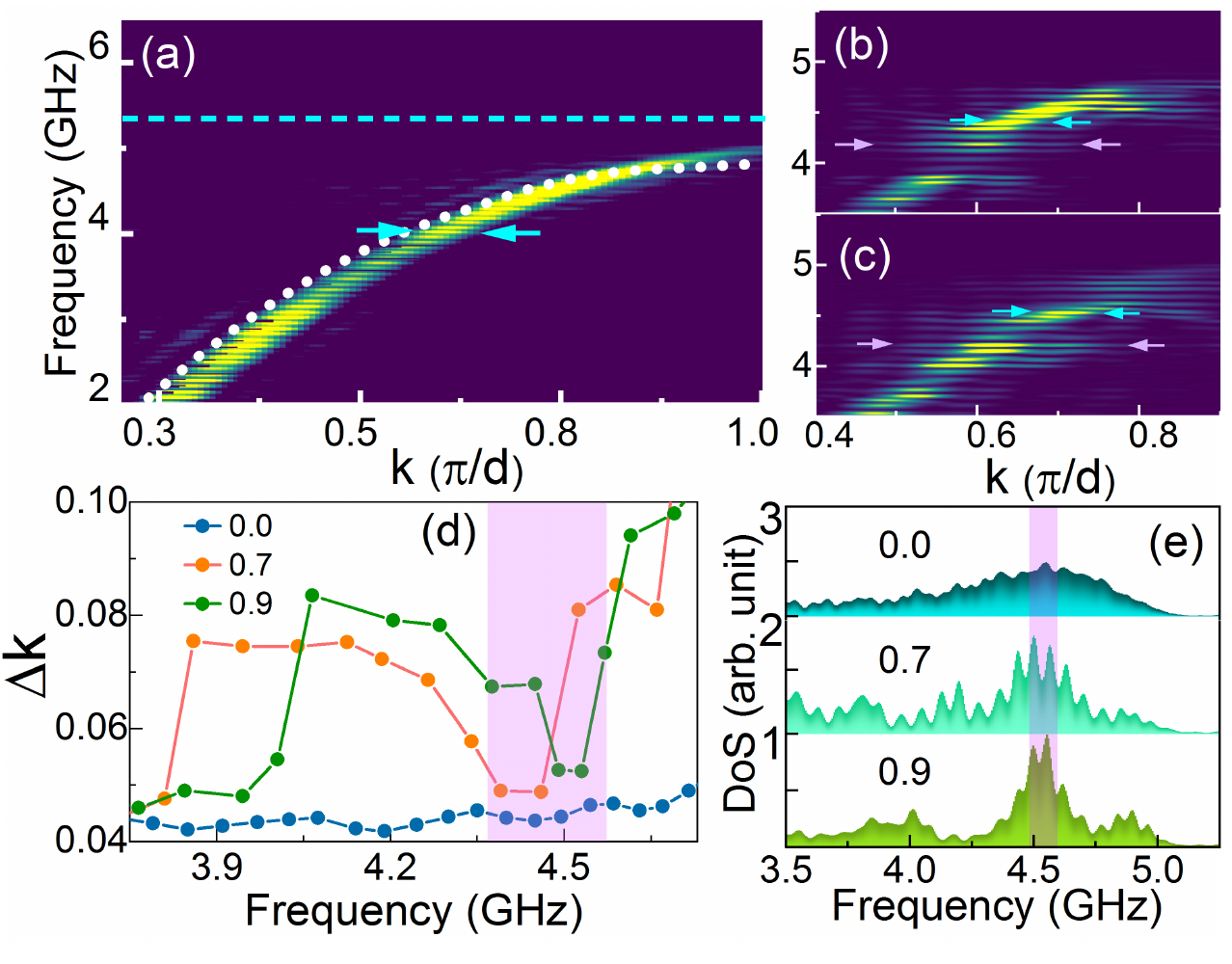}
\caption{Experimentally measured dispersion for the finite SSPP structures. (a) Dashed cyan line Non-dispersive resonant mode (5.35~GHz) for an infinite sample. Dotted white curve: Hybrid mode at the bandedge $\sim 4.77$GHz. Measured experimental results (Viridis colormap) show the hybrid band in excellent agreement with the predicted band. Arrows demarcate the FWHM of the $k$-space peaks. Highly disordered structures: (b) $\delta=0.7$ and (c) $\delta=0.9$. (d) $\Delta k$ as a function of frequency, measured at the three values of disorder. The band-edge region (vertical purple band) shows a drop in $\Delta k$. (e) Measured DoS as a function of frequency. The band-edge region (purple band) shows an enhanced DoS, a consequence of frequency pinning in hybrid systems.}\label{fig:dispersion}
\end{figure}

The hybridization occurring in this system(see Supplementary information S2) was experimentally verified by measuring the dispersion diagram of the finite-sized periodic and disordered structures, and comparing it with the expected resonance characteristics. Fig.\ref{fig:dispersion}(a), \ref{fig:dispersion}(b) and \ref{fig:dispersion}(c) show the said dispersion. The white dotted curve in \ref{fig:dispersion}(a) illustrates the dispersion behavior computed using the eigenmode solver of COMSOL Multiphysics after applying periodic boundary conditions. The band arises from the hybridisation of SPP dispersion with the resonant modes of the individual cavities. The latter are non-dispersive, and are shown as a horizontal cyan dashed line, whose resonant frequency was calculated to be 5.35~GHz. The inset shows the corresponding resonant mode for a single resonator. The experimental modes are in excellent agreement with the computed dispersion, confirming successful excitation of spoof SPP's formed over the structured metal surface. The band-edge lay at 4.77~GHz. The cyan arrows indicate the width $\Delta k$ at any frequency, which is inversely proportional to the spatial extent of the mode at the frequency. Fig.\ref{fig:dispersion}(b) and \ref{fig:dispersion}(c) show the measured dispersion of two particular configurations at high disorder, $\delta = 0.7$ and $0.9$ respectively. Disorder clearly broadens the $\Delta k$ at a few frequencies, as labeled by the arrows, reflecting the spatial localization of modes. The violet arrows represent localized modes as inferred from the spatial studies (discussed further), while the cyan arrows indicate persisting extended states in the structure even under strong disorder. The co-existence of localized and extended states is revealed in subplot \ref{fig:dispersion}(d), which illustrates the systematic variation of $\Delta k$ with frequency. Blue circles represent the periodic structure, with narrow and almost uniform $\Delta k$ across the frequency axis.  Orange  and green circles ($\delta = 0.7 ~\&~ 0.9$, respectively) illustrate a rise in the $\Delta k$ at frequencies where localization sets in. Both the curves show a fall in the shaded frequency region, where the extended  states manifest.

Theoretically, it is expected that these extended states lie in a miniband formed by the coalescence of eigenvalues close to the bandedge.
Interestingly, we note that a similar phenomenon called ``eigenvalue condensation" was theoretically observed for strong randomness in imaginary potentials\cite{tzortzakakis2021transport}. In the experimental data, such a miniband can be directly inferred from the increased DoS shown in the same frequency (shaded) region, as illustrated in (e). In the weak disorder limit($\delta=0.0,~0.3$), the DoS shows a typical behavior where it consistently increases upto the bandedge. The miniband manifests at $\delta = 0.7$, and is strongly conspicuous at $\delta = 0.9$.

\begin{figure}[h]
\centering
\includegraphics[width=1\linewidth]{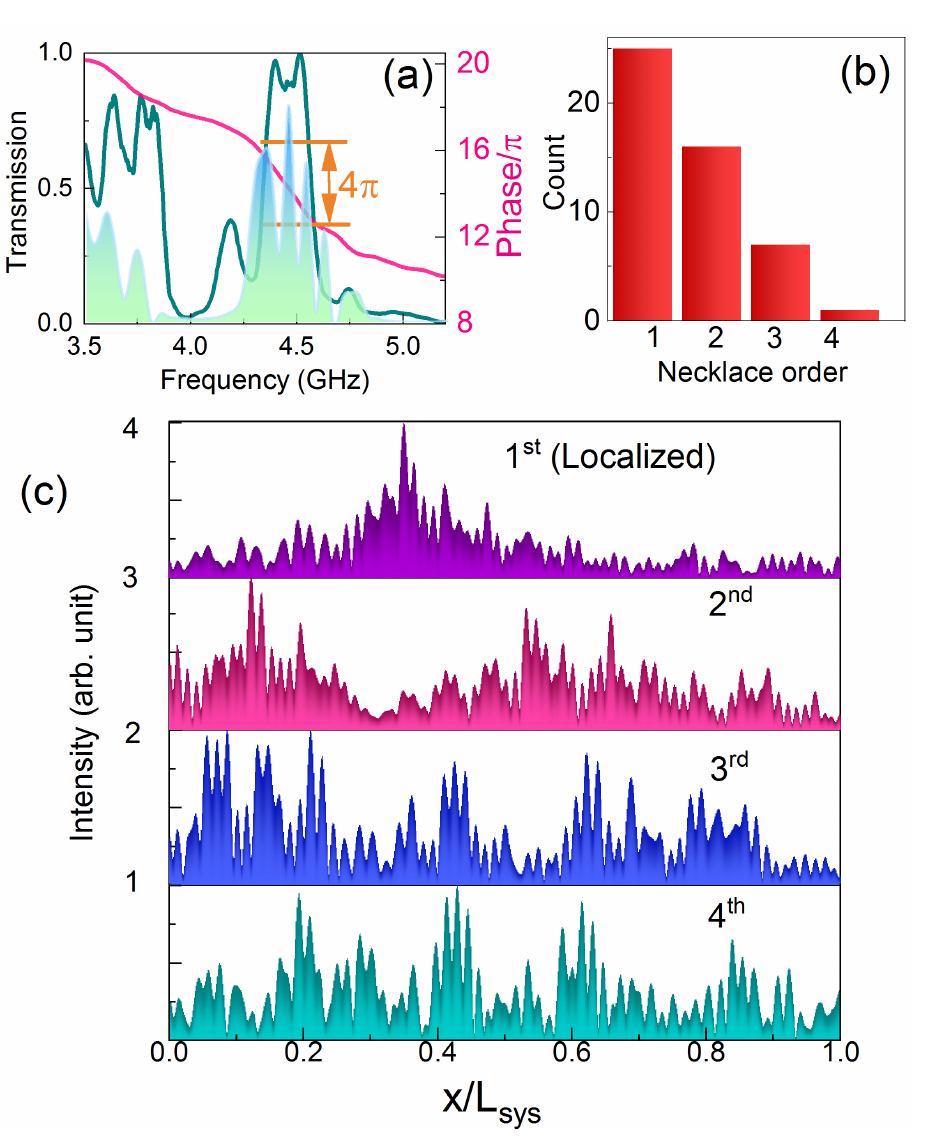}
\caption{Measurement of phase and necklace state order in samples with high disorder. (a) Experimentally measured transmission spectrum (green solid line) and the corresponding extracted phase (pink solid line) for a single configuration at high disorder. A drop in phase by a magnitude of $4\pi$ endorses a $4^{th}$ order necklace state. The simulated transmission spectrum (shaded green) is in excellent agreement with the experimental measurements and indeed reveal four resonances in the necklace (See Supplementary Information S4). (c) shows the necklace order $m$ distribution measured over 49 modes. (c) Spatial intensity distributions measured for different orders of necklace states. See Supplementary information S6. }\label{fig:phase}
\end{figure}

In the hybrid plasmonic system under high disorder, the eigenvalues accumulate in the vicinity of the hybridization frequency. Our simulations detected this behavior in the eigenvalues, as discussed in the Supplementary information S3. The non-Hermiticity lifts the orthogonality of the eigenmodes and enables energy transport between them. In this scenario, whenever multiple localized eigenmodes overlap even minimally in the tails, the spoof polariton can be transmitted through the chain of eigenmodes, that is, the necklace states. Such necklace states,  essentially coupled localized resonances, can be analyzed by observing the phase jumps in the constituent phase profile \cite{bertolotti2006wave} measured from complex field at the output end of the samples. Every component resonance effects a phase shift of $\pi$.  Fig.\ref{fig:phase} discusses the phase behavior of a particular necklace state observed in a sample with high disorder. The green curve depicts the measured transmission spectrum (left Y-axis). The peak at around 3.7~GHz represents the photonic bands wherein localization hasn't yet set in. The drop in transmission at $\sim4~$GHz represents incipient localization. Next, a broad anomalous peak is seen in the vicinity of the bandedge at around $4.5~$GHz which represents the miniband formed due to the coalescence of eigenvalues. The peak at $\sim$4.2~GHz is not a consequence of the structure, as ascertained from various configurations. The phase, as extracted from the complex $E_z$ measurement, is shown in the pink curve (right Y-axis). A clear jump of 4$\pi$ is observed over the frequency range pertinent to the miniband. Corresponding simulations for this particular disorder configuration provided the theoretical transmission spectrum shown by the overlapping shaded green region. Eigenvalue analysis of this spectrum clearly revealed four constituent coupled resonances creating the necklace state. See Supplementary information S4 for details. Over the measured 49 states in 21 samples, the distribution of localized modes and necklace states is shown in the bar graph in \ref{fig:phase}(b). The measured phase jump of 4$\pi$ endorses this coupling. Several such necklace states of various orders were observed, and a few representative states are shown in Fig.\ref{fig:phase}(c). The top panel shows an order-1, i.e, essentially a conventional Anderson localized mode, with a localization length $\xi/L_{sys} = 0.3$. The lower panels exhibit necklace states with orders 2, 3 and 4 respectively, as characterized by the measured phase jumps. As seen, almost 50\% localized states couple to form necklace states of some order. This fact distinguishes the hybrid non-Hermitian system from conventional systems wherein the necklace states are statistically rare and do not contribute to average transport. In contrast, the high probability of necklace states in this system induces anomalous transport. 

\begin{figure}[h]
\centering
\includegraphics[width=1\linewidth]{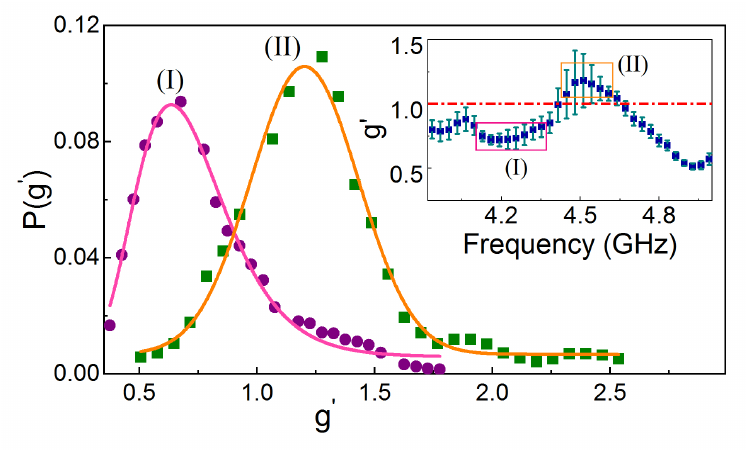}
\caption{Effect of disorder on generalized conductance. Inset: Measured $\langle g' \rangle$ from intensity distributions averaged over 21 configurations as a function of frequency. The dashed line at $ \langle g'\rangle=1$ demarcates localization regime. Main plot: $P(g')$ measured at two regions as marked by pink and orange rectangles in (a): (I)Localized region ($\langle g'\rangle<1$): Long-tailed distribution, best fit with a log-normal function (solid line) with mean $\langle g' \rangle$ = 0.6 and width $\sigma_{g} = 0.07$. (II) Anomalous transmission regime ($\langle g' \rangle \geq 1$): The distribution exhibits a Gaussian profile with mean $\langle g' \rangle = 1.2$ and width $\sigma_{g} = 0.2$. }\label{fig:conductance}
\end{figure} 

The physics of disordered systems is completely described by the behavior of conductance. Therefore, in order to demarcate the novel regime of transport uncovered in this work, we discuss the conductance of the samples in the presence of the necklace states. Figure 5 depicts the behaviour of generalized conductance $g'$, a parameter derived from intensity statistics that can faithfully characterize localization even in dissipative systems\cite{sebbah2006extended}. Since our system is inherently non-Hermitian, $g'$ provides the sole unambiguous characterisation of transport, which is illustrated in subplot (a). The $g'$ is provided by $2/3$var$(I)$, where $I$ is either the normalized total transmission, or in-plane spatial intensity in a one-dimensional sample\cite{chabanov2000statistical,joshi2020reduction}. We extracted the generalized conductance $\langle g' \rangle$ over 21 highly disordered configurations. Fig.\ref{fig:conductance}(a) shows the average generalized conductance $\langle g' \rangle$ vs frequency, with $\langle g' \rangle =1$ delimiting the onset of the Anderson localized regime. In the vicinity of $4.5$~GHz, the $\langle g' \rangle$ is clearly seen to rise above 1, endorsing the anomalous transport. The error bars indicate the standard deviation of the  $ g' $ values indicating that the majority of the samples were close to, or in the anomalous transport regime. For lower frequencies, $\langle g'\rangle < 1$) for the localized regime.  The distribution of conductances $P(g')$ follows a stipulated behavior in disordered systems. The rectangles (I) \& (II) in \ref{fig:conductance}(a)  demarcate the regions over which $g'$ values were chosen to create the $P(g')$, shown in Fig.\ref{fig:conductance}(b). In the localized regime (magenta dots from rectangle (I)), the distribution is asymmetric and long-tailed, and nicely fit by a log-normal function with mean $0.6$, as seen from solid the magenta curve. That conductances in the localized regime are distributed log-normally is a well-established fact, also confirmed experimentally\cite{kumar2020discrepant, mondal2019optical}. On the other hand, in the anomalous transport regime(ATR) (green squares, data from rectangle (II)), the distribution is symmetric albeit with a longer tail.  The data are excellently fit by a Gaussian distribution, as is expected in the metallic regime. The mean $\langle g' \rangle  = 1.2$ from the Gaussian fit.

These data lie at the boundary of localization-delocalization. For this regime, we apply the theory established in \cite{muttalib2003conductance}, where the authors theoretically examine the conductance distributions in one-dimensional disordered systems. In a particular situation where localized and extended states co-exist, the $P(g')$ for the metallic states is shown to follow a Gaussian function $P(g')\propto\exp({{-\frac{15}{2} {(g'-\langle g' \rangle)^2})}}$. Clearly, the experimental data in Fig~5(b) shows an excellent fit to this equation, which provides a value of $\langle g' \rangle = 1.16$, whereas the experimental value was measured to be 1.2, once again in excellent agreement. These data conclusively endorse the existence of the system in a properly conducting regime, despite insulating disorder. The insulating nature of disorder is certified by the fact that, at frequencies away from the bandedge where necklaces are not yet forged, the conductances are consistently sub-unity. The probability distributions ($P(I/\bar{I})$)of normalized intensity measured at localized and necklace state regimes is provided in Supplementary information S5. 

In conclusion, we have successfully uncovered an anomalous transport regime in hybrid, non-Hermitian systems. One-dimensional structures supporting spoof surface plasmon polaritons were investigated for the same. Conventional Anderson localization leading to arrested transport was observed at moderate disorders. However, at high disorder, an  enhancement in transmission is facilitated by formation of necklace states. The hitherto reported necklace states in localizing systems are rare and do not contribute to the average transport. In contrast, in the hybrid non-Hermitian system, the necklace states manifest persistently for all configurations at high disorder. The anomalous transmission manifests in the vicinity of the hybridization frequency, where frequency-pinning of modes occurs due to constrained migration. The resulting emergent mini-band has been experimentally demonstrated in our experiments. Direct mode-mapping using a near-field spatial probe allowed us to image the necklace states, as well as identify the order thereof via phase jumps. The latter provided a direct measure of the component localized modes within the state, which was in excellent agreement with our numerical calculations on the various configurations. The transport regime is labeled via the measurement of generalized conductance $\langle g' \rangle$, which was consistently above 1, endorsing high transmission even under localizing conditions. Comparing the distribution of $g'$ with theoretical expressions for one-dimensional disordered systems close to the localization transition, we found that our system was indeed in  a `metallic' state despite the strong disorder.

\begin{acknowledgments}
We thank M. Balasubrahmaniyam and Krishna Chandra Joshi for useful discussions and helpful comments. We thank Meghan Patankar for his help in sample making. We acknowledge funding from the Department of Atomic Energy, Government of India (12-R\&D-TFR-5.02-0200). S. Mujumdar acknowledges the Swarnajayanti Fellowship from the Department of Science and Technology, Government of India.
\end{acknowledgments}

\end{document}